\def\editmode{0}

\def\bibfilenames{refs,bibman_refs}

\def\singlenarrowcol{0}
\if\singlenarrowcol0
    \documentclass[conference]{IEEEtran}

    \IEEEoverridecommandlockouts

\else
    \documentclass[onecolumn]{IEEEtran}
    \usepackage[pass]{geometry}
    \newgeometry{textwidth=252pt, textheight=672pt}
    
\fi

\usepackage{include_v7}

\usepackage{hyperref}

\usepackage{import}

\def\journal{0}

\usepackage{environ}

\if\journal0    
    \NewEnviron{journalonly}{%
    }
    \NewEnviron{conferenceonly}{%
    \BODY
    }
\else
    \if\journal1
        \NewEnviron{journalonly}{%
        \BODY
        }   
        \NewEnviron{conferenceonly}{%
        }  
    \else
        \if\journal2

        \fi
    \fi
\fi

\def\intermediatesteps{0}
\if\intermediatesteps1
    \NewEnviron{intermed}{%
        \color{purple}
        \BODY
    }

\else
    \NewEnviron{intermed}{%
    }

\fi

\newcommand{\Area}{\hc{\mathcal{X}}}
\newcommand{\Loc}{\hc{\bm x}} 
\newcommand{\TxSig}{\hc{s}} 
\newcommand{\RxSig}{\hc{r}} 
\newcommand{\Cha}{\hc{h}} 
\newcommand{\Noi}{\hc{z}} 
\newcommand{\NumFea}{\hc{N}} 
\newcommand{\IndFea}{\hc{n}} 
\newcommand{\NotFea}[1]{_{#1}} 
\newcommand{\Rss}{\hc{f}} 
\newcommand{\EstRss}{\hc{\hat f}} 
\renewcommand{\expected}{\hc{\mathbb{E}}} 
\newcommand{\NumSam}{\hc{K}} 
\newcommand{\SetSam}{\hc{\mathcal{K}}} 
\newcommand{\EstVecRss}{\hc{\hbm f}} 
\newcommand{\NumFra}{\hc{J}} 
\newcommand{\IndFra}{\hc{j}} 
\newcommand{\NotFra}[1]{\hc{[#1]}} 
\newcommand{\SetIndLeg}{\hc{\mathcal{J}_\text{L}}} 
\newcommand{\SetIndAtt}{\hc{\mathcal{J}_\text{A}}} 
\newcommand{\NumLeg}{\NumFra_{L}} 
\newcommand{\NumAtt}{\NumFra_{A}} 
\newcommand{\EstMatRss}{\hc{\hbm F}}
\renewcommand{\hz}{\hc{\mathcal{H}}_0}
\renewcommand{\ho}{\hc{\mathcal{H}}_1}
\newcommand{\NumDat}{\hc{M}} 
\newcommand{\NotDat}[1]{_{#1}} 
\newcommand{\IndDat}{{\hc{m}}} 
\newcommand{\NotEst}[1]{^{(#1)}} %
\newcommand{\IndEst}{{\hc{i}}} 
\newcommand{\HzPCD}{\hc{\hz^\text{PCD}}} %
\newcommand{\HoPCD}{\hc{\ho^\text{PCD}}} %
\newcommand{\NumPai}{\hc{P}} 
\newcommand{\IndPai}{{\hc{p}}} 
\newcommand{\DatSam}{\hc{\mathcal{D}_\text{s}}} 
\newcommand{\DatDif}{\hc{\mathcal{D}_\text{d}}} 
\newcommand{\DecPCD}{\hc{d^\text{PCD}}} 
\newcommand{\Val}{\hc{_\text{val}}} 
\newcommand{\Tra}{\hc{_\text{tr}}} 

\newcommand{\samp}{\hc{\epsilon}} 

\newcommand{\AuxNetWei}{\hc{\tilde T}^{\text{PCD}}} %

\newcommand{\VecRss}{\bm\Rss} 

\newcommand{\Gra}{\hc{\mathcal{G}}} 
\newcommand{\NodeInd}{\hc{\nu}} 
\newcommand{\Nei}{\hc{\mathcal{N}}} 
\newcommand{\LayerInd}{\hc{l}} 
\newcommand{\NumLayer}{\hc{L}} 

\newcommand{\LocLin}{\hc{\tbm x}} 
\newcommand{\Spe}{\hc{v}} 
\newcommand{\VecDir}{\hc{\bm d}} 

\newcommand{\IndSam}{\hc{k}} 

\newcommand{\SamPer}{\hc{T}} 

\newcommand{\NumEst}{\hc{I}} 

\newcommand{\Dat}{\hc{\mathcal{D}}} 

\newcommand{\RatFra}{\hc{R}_\text{J}} 

\newcommand{\HzSD}{\hc{\hz}} %
\newcommand{\HoSD}{\hc{\ho}} %

\newcommand{\Net}{\hc{T}^{\text{PCD}}} 
\newcommand{\NetWei}{\Net} 

\newcommand{\Reg}{\hc{\mathcal{R}}} 
\newcommandoa{\IndReg}{\hc{r}}{\hc{r}[#1]} 

\allowdisplaybreaks
\begin{document}

\title{Spoofing Detection in the Physical Layer with Graph Neural Networks\thanks{This research has been funded in part by the Research Council of Norway under IKTPLUSS grant 311994.}\thanks{The code and pre-trained models are available at \url{https://github.com/uiano/gnn_spoofing_detection}.}
}

\author{
    \IEEEauthorblockN{Tien Ngoc Ha}
    \IEEEauthorblockA{\textit{
            Department of ICT},
        \textit{University of Agder}\\
        Grimstad, Norway \\
        tien.n.ha@uia.no}

    \and
    \IEEEauthorblockN{Daniel Romero}
    \IEEEauthorblockA{\textit{
            Department of ICT},
        \textit{University of Agder}\\
        Grimstad, Norway \\
        daniel.romero@uia.no}
}


\maketitle

\begin{abstract}
    In a spoofing attack, a malicious actor impersonates a legitimate user to access
    or manipulate data without authorization. The vulnerability of cryptographic security mechanisms to compromised user credentials motivates spoofing attack detection in the physical layer, which  traditionally relied on  channel features, such as
    the received signal strength (RSS) measured by spatially distributed
    receivers or access points. However, existing methods cannot effectively cope with the dynamic nature of channels, which change over time as a
    result of user mobility and other factors. To address this limitation,
    this work builds upon the intuition that the temporal pattern of changes in RSS features can be used to detect the presence of concurrent transmissions from multiple (possibly changing) locations, which in turn indicates the existence of an attack. Since a localization-based approach would require costly data collection and would suffer from low spatial resolution due to multipath, the proposed algorithm employs a deep neural network to construct a graph embedding of a sequence of  RSS features that reflects changes in the propagation conditions. A graph neural network then classifies these embeddings to detect spoofing
    attacks. The
    effectiveness and robustness of the proposed
    scheme are corroborated  by experiments with real-data. 
\end{abstract}


\section{Introduction}
\label{sec:intro}
\begin{bullets}

    \blt[Motivation]The prevalence of wireless communications has engendered a
    panoply of security threats, including the unauthorized interception of
    private data, disruptions to remote services, and user impersonation. Among
    these pernicious threats, spoofing attacks pose a particularly troublesome
    challenge since, in these attacks, malevolent actors intercept and manipulate
    data originally intended for legitimate users~\cite{kolias2015intrusion,
        vanhoef2016mac, martin2019handoff, tippenhauer2011requirements}. The
    detection and mitigation of these attacks are pivotal for data security.
    Although cryptographic techniques have traditionally been employed across
    various communication layers to fortify security, the potential access of
    attackers to the credentials of legitimate users introduces a serious
    vulnerability. Consequently, the research community has increasingly focused
    on detecting spoofing attacks in the physical layer.



    \blt[Literature]
    \begin{bullets}
        \blt[Review]
        \begin{bullets}
            \blt[Overview]For instance,
            \cite{brik2008wireless,givehchian2022evaluating,liu2019real,vo2016fingerprinting} leverage transmitter hardware imperfections, such as
            carrier frequency offset (CFO), in-phase and quadrature (I/Q) offset, and I/Q imbalance,  to verify user identity. Regrettably, these methodologies
            necessitate knowledge of the communication
            protocol and may prove ineffective in the face of environmental
            changes, such as fluctuations in temperature
            \cite{givehchian2022evaluating}.
            \blt[AoA, TDoA, SNR]These constraints are somehow mitigated
            in
            \cite{xiong2010secureangle,xiong2013securearray,shi2016robust},
            which rely on angle of arrival (AoA) and time difference of arrival (TDoA) features,
            and in \cite{wang2020machine}, where a neural network is
            trained using signal-to-noise ratio (SNR) traces. Nonetheless, these
            approaches still demand synchronization and/or knowledge of the
            communication protocol.
            \blt[RSS]In contrast, techniques reliant on received signal strength (RSS) measurements do not
            require knowledge of the communication protocol or signal decoding,
            thus significantly augmenting their generality and applicability for
            detecting spoofing attempts
            \cite{chen2007detecting,yang2012detection,xiao2016phy,zeng2011identity,alotaibi2016new}.
            The predominant approach in this context involves applying
            clustering primitives to RSS measurements collected by multiple receivers, such as the access points of a WiFi network
            \cite{chen2007detecting,hoang2018soft,sobehy2020csi}. By exploiting
            the dependence of RSS signatures on the transmitter
            locations, an attack is detected if transmissions with the same user
            identifier are found to originate at different locations.
        \end{bullets}%
        \blt[Drawbacks] Consequently, this approach results in false alarms when the channel
        conditions change, as for example when a legitimate user moves.

    \end{bullets}

    \blt[Contributions]
    \begin{bullets}
        \blt[Key realization]To remedy this limitation, the key realization in
        this work is that it is possible to tell spoofing from motion and other
        effects by analyzing the temporal changes in RSS features. To illustrate this
        idea, consider a network that sequentially receives frames from
        locations denoted as A, B, C, and D. If all these locations are
        distinct, it is natural to ascribe these variations to  the movement of the legitimate user.
        In contrast, if the received transmissions alternate between points A
        and B in a pattern such as A, B, A, B, A, B, etc., it is more likely
        that one user is transmitting from location A and another from location
        B, which indicates the presence of an attack.
        \blt[novelty]To the best of our knowledge, the work at hand is the first to exploit this kind of information.

        \blt[Overview]To this end, this paper introduces a spoofing
        attack detection scheme where a graph embedding is constructed to capture the pattern of changes in RSS features over a sequence of frames. Then, a \emph{graph neural network} (GNN)  classifies such graph embeddings as either corresponding to an attack or to legitimate user activity, which may include user movement.
        The graph is constructed by utilizing a \emph{position-change detector} (PCD) that determines whether a given pair of frames was transmitted from different locations. Since changes in the RSS measurements corresponding to different frames may be caused either by the movement of the transmitter or by the variability due to the finite number of samples used in the computation of these measurements, the PCD is designed as a deep neural network that detects position changes by implicitly learning the distribution of  RSS estimates from signal samples.
        \blt[deployment]The proposed scheme can be readily deployed due to the simplicity of the procedure for collecting the required data set. Specifically, RSS features must be collected at different locations but those locations need not be recorded.

    \end{bullets}

    \blt[Paper structure] The rest of the paper is structured as follows.
    Sec.~\ref{sec:problem} formulates  the problem. Sec.~\ref{sec:proposed} presents the proposed spoofing detection scheme. Sec.~\ref{sec:performance} presents an extensive performance evaluation using
    real data. Finally, Sec.~\ref{sec:conclusion} concludes the paper.

\end{bullets}

\section{Problem Formulation}
\label{sec:problem}

\cmt{model}
\begin{bullets}

    \blt[Space] Let $\Area \subset \rfield^3$ comprise the  coordinates of
    all points in the  spatial region of interest, where  both
    legitimate users and attackers are located.
    \blt[Feature vectors]
    \begin{bullets}%
        \blt[Received signal as a function of position and time]%
        \begin{bullets}%
            \blt[tx] A transmitter  at $\Loc\in\Area$, which can be the
            legitimate user or an attacker, sends a signal $\TxSig(t)$, where
            $t$ denotes time. This signal, modeled as an unknown wide-sense
            stationary stochastic process, \blt[rx] is received by $\NumFea$
            receivers, such as the access points or base stations of a wireless network. Let $\Cha\NotFea{\IndFea}(\Loc,t)$ denote the  impulse
            response of the channel between the transmitter and the $\IndFea$th
            receiver, which is assumed to be time-invariant over the duration of a frame. The received signal at the $\IndFea$th receiver
            is given by
            \begin{align}
                \RxSig\NotFea{\IndFea}(\Loc,t) = \Cha\NotFea{\IndFea}(\Loc,t) \ast \TxSig(t) + \Noi\NotFea{\IndFea}(t),
            \end{align}
            where  $\ast$ denotes
            convolution and $\Noi\NotFea{\IndFea}(t)$ is  additive white
            Gaussian noise (AWGN) with variance $\sigma^2$ and
            independent of $\TxSig(t)$.
            \blt[rss]Thus, one can define the \emph{received signal strength}
            (RSS) $\Rss\NotFea{\IndFea}(\Loc) \define \expected{|
                    \RxSig\NotFea{\IndFea}(\Loc,t) |^2},$ where $\expected$ denotes
            expectation.
        \end{bullets}

        \blt[RSS estimates] To estimate the RSS of a frame received by  the $\IndFea$th receiver, consider a set of $\NumSam$
        samples
        $\SetSam\NotFea{\IndFea} \define \{
            \RxSig\NotFea{\IndFea}(\Loc,t+\IndSam\SamPer) \}_{\IndSam=1}^{\NumSam}$,
        where $\SamPer$ is the sampling period and $t$ is the time when the frame begins. The RSS can be estimated as
        \begin{align}
            \label{eq:rssestimator}
            \EstRss\NotFea{\IndFea}(\Loc) \define \frac{1}{\NumSam}
            \sum_{\IndSam=1}^{\NumSam} |
            \RxSig\NotFea{\IndFea}(\Loc,t+\IndSam\SamPer) |^2.
        \end{align}
        If
        $\RxSig\NotFea{\IndFea}(\Loc,\IndSam \SamPer)$ is ergodic for each
        $\Loc$,  it follows that $\EstRss\NotFea{\IndFea}(\Loc)$ converges to
        $\Rss\NotFea{\IndFea}(\Loc)$ as $\NumSam\rightarrow \infty$.

        \blt[RSS vector]For notational convenience, the
        RSS values estimated by all receivers are collected into a \emph{feature vector}, generically represented by
        $\EstVecRss(\Loc) \define [
                \EstRss\NotFea{1}(\Loc), \ldots, \EstRss\NotFea{\NumFea}(\Loc)
            ]\transpose$.
        \blt[multiple estimates]Note that since $\NumSam$ is finite,
        measuring the RSS $\NumEst$ times for location $\Loc$ yields $\NumEst$
        different estimates of $\VecRss(\Loc)\define[
                \Rss\NotFea{1}(\Loc),
                \ldots,
                \Rss\NotFea{\NumFea}(\Loc)
            ]\transpose$. These  estimates will be denoted as
        $\EstVecRss\NotEst{\IndEst}(\Loc),
            \IndEst=1,\ldots,\NumEst$.


    \end{bullets}

    \blt[Frames and user indices]To introduce the notation for frame sequences,
    \begin{bullets}%
        \blt[Frame definition]let $\Loc\NotFra{\IndFra}$ denote the location of the user that transmits the $\IndFra$th frame at the moment of transmitting that frame and let $\EstVecRss\NotFra{\IndFra}
            \define \EstVecRss(\Loc\NotFra{\IndFra})$.
        \blt[Frame sequence]The feature vectors corresponding to  a  sequence
        of  $\NumFra$ frames are collected into matrix $\EstMatRss \define [ \EstVecRss\NotFra{1},
                \ldots, \EstVecRss\NotFra{\NumFra} ]$.
        \blt[User indices]Out of these
        $\NumFra$ frames, $\NumLeg$  belong to the legitimate user and $\NumAtt$ to the
        attacker, where $\NumFra = \NumLeg + \NumAtt$.
        \begin{bullets}
            \blt[Legitimate]The
            set of indices of the frames belonging to the legitimate user is represented by  $\SetIndLeg \subset \{1,\ldots,\NumLeg\}$
            \blt[Attacker]whereas the set of indices of the frames belonging to the attacker is represented by  $\SetIndAtt \subset \{1,\ldots,\NumAtt\}$.

        \end{bullets}
    \end{bullets}
\end{bullets}

\cmt{Problem formulation}
\begin{bullets}
    \blt[Given]Given  $\EstMatRss$,
    \blt[Decisions/Hypotheses]the problem is to decide between the following
    hypotheses:
    \begin{align}
        \label{eq:problem}
        \left\{\begin{matrix}
                   \hz: \SetIndAtt = \emptyset \\
                   \ho: \SetIndAtt \neq \emptyset.
               \end{matrix}\right.
    \end{align}
    \blt[Dataset]To this end, a dataset comprising the feature vectors
    $\Dat\define\{\EstVecRss\NotEst{\IndEst}(\Loc\NotDat\IndDat), \IndDat=1,\ldots,\NumDat,
        \IndEst=1,\ldots,\NumEst \}$ is given, where $\NumDat$ is the number of distinct measurement locations, i.e.
    $\Loc\NotDat\IndDat \ \neq \Loc\NotDat{\IndDat'}~\forall \IndDat \neq
        \IndDat'$.
\end{bullets}

\section{Spoofing Detection from RSS Features}
\label{sec:proposed}


\begin{figure}[t]
    \centering
    \includegraphics[width=1.\columnwidth]{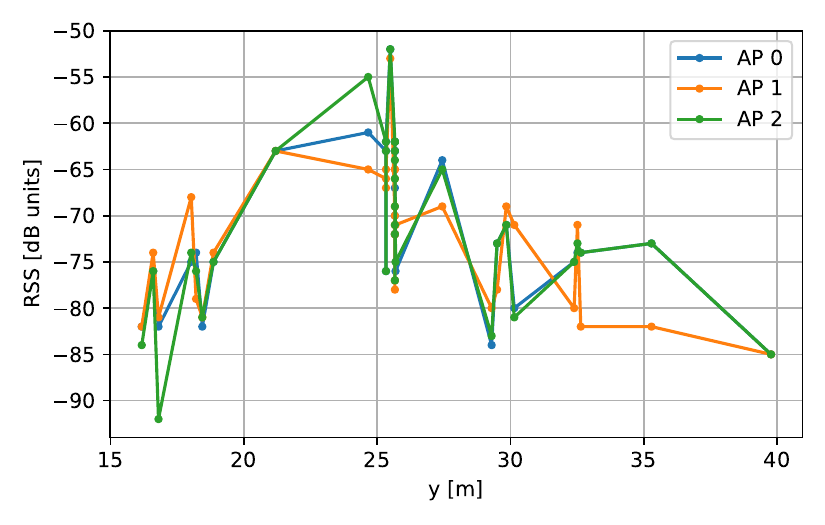}
    \caption{RSS measurements along a line vs. their y-coordinate. It is observed that small variations in the y-coordinate often result in larger RSS changes than large variations in the y-coordinate. For example, the difference between $y=24$ and $y=29$ is around 24 dB, whereas the difference between $y=16$ and $y=40$ is less than 5 dB.
        This is caused mainly by multipath and suggests that accurately estimating the position from RSS measurements is not generally possible.}
    \label{fig:fading}
\end{figure}

\begin{bullets}%
    \blt[assumption on high frame rate]%
    \begin{bullets}%
        \blt[motivation]
        Evidently, if the time between consecutive frames in $\EstMatRss$ is too long, then
        the vectors $\EstVecRss\NotFra\IndFra$ may
        originate at highly distant
        locations due to user movement,
        even in the absence of attacks. As a result, $\EstVecRss\NotFra\IndFra$ may be  highly different from $\EstVecRss\NotFra{\IndFra-1}$ and
        $\EstVecRss\NotFra{\IndFra+1}$. Since this would also be the case in the presence of an attack,
        solving problem \eqref{eq:problem} becomes challenging as the
        distributions of $\EstMatRss$ under both
        hypotheses are highly similar. Therefore, it becomes imperative to introduce the following
        assumption:

        \blt[Statement] \emph{Assumption 1: The frame rate is high relative
            to the speed of the users.}

        \blt[clarification]In other words, if $\EstVecRss\NotFra{\IndFra}$ and $\EstVecRss\NotFra{\IndFra+1}$ correspond to the same user, they will be reasonably similar.

    \end{bullets}%

    \blt[Motivation of PCD \ra why not loc] In view  of Assumption 1, one could consider a strategy
    to tackle problem \eqref{eq:problem} where the locations
    $\Loc\NotFra\IndFra$ are first estimated based on  $\EstVecRss\NotFra\IndFra$, $\IndFra=1,\ldots,\NumFra$, and an attack is declared if multiple transmissions are concurrently received from distant (possibly moving) users. However, this approach is not viable first because the data given in the problem formulation of Sec.~\ref{sec:problem} does not allow a reasonably accurate localization of the transmitters. Indeed,
    if the locations $\Loc\NotFra\IndFra$ associated with the vectors in
    $\Dat$ were given, one could attempt to estimate the locations associated with the frames in $\EstMatRss$, for instance via fingerprinting-based localization~\cite{lohan2017crowdsourced, barsocchi2016multisource}.  However, the error of such approaches is typically in the order of 10 m in indoor environments~(see e.g.~\cite{lohan2017crowdsourced}), which would hinder  detecting attacks where the attacker is relatively near the legitimate user. Fig.~\ref{fig:fading} illustrates why this is the case.   Besides, collecting a data set where the positions of the measurement locations need to be recorded is highly costly since it would generally require the deployment of an auxiliary localization system, the use a mobile robot, or to manually measure the spatial coordinates of all measurement locations.


    \blt[Proposed approach] For this reason, the proposed scheme does not attempt to estimate the transmitter locations. Instead, it exploits the pattern of dissimilarities between the feature vectors $\EstVecRss\NotFra\IndFra$. This is accomplished in two steps: First, each pair of vectors $(\EstVecRss\NotFra\IndFra,\EstVecRss\NotFra{\IndFra'})$ is compared as described in
    Sec.~\ref{sec:poschange}. Given these comparisons, a decision is made on the presence of an attack based on a graph embedding, as described in Sec.~\ref{sec:gnn_detection}.

\end{bullets}

\subsection{Position-change Detection}
\label{sec:poschange}

\begin{bullets}%
    \blt[Overview] This section presents a PCD, which is a detector that determines whether two given frames where transmitted from  the same location.
    \blt[Problem formulation]%
    \begin{bullets}%
        \blt[Given]Specifically, given two feature vectors $\EstVecRss\NotFra{\IndFra}$ and
        $\EstVecRss\NotFra{\IndFra'}$ respectively corresponding to (possibly equal) locations
        $\Loc\NotFra{\IndFra}$ and $\Loc\NotFra{\IndFra'}$,
        \blt[Decisions/Hypotheses]the goal is to distinguish between the following hypotheses:
        \begin{align}
            \label{eq:pcdprob}
            \left\{\begin{matrix}
                       \HzPCD: \Loc\NotFra{\IndFra} = \Loc\NotFra{\IndFra'} \\
                       \HoPCD: \Loc\NotFra{\IndFra} \neq \Loc\NotFra{\IndFra'}.
                   \end{matrix}\right.
        \end{align}
        A PCD is a function that maps a pair of feature vectors to a hypothesis, i.e.,
        $\DecPCD: \rfield^{\NumFea} \times \rfield^{\NumFea} \rightarrow \{\HzPCD,
            \HoPCD\}$.
    \end{bullets}

    \blt[why DNN]To properly address \eqref{eq:pcdprob}, it is useful to consider the components behind the dissimilarity between   $\EstVecRss\NotFra{\IndFra}$ and
    $\EstVecRss\NotFra{\IndFra'}$.
    \begin{bullets}%
        \blt[causes for changes]
        \begin{bullets}%
            \blt[finite $\NumSam$]First, two
            feature vectors $\EstVecRss\NotFra{\IndFra}$ and
            $\EstVecRss\NotFra{\IndFra'}$ are naturally different because $\NumSam$ is finite,
            even when $\Loc\NotFra{\IndFra}=\Loc\NotFra{\IndFra'}$.
            \blt[movement]Second, $\EstVecRss\NotFra{\IndFra}$ and
            $\EstVecRss\NotFra{\IndFra'}$ will be different when $\Loc\NotFra{\IndFra}\neq\Loc\NotFra{\IndFra'}$ because of the different propagation phenomena undergone by the signals propagating from either location to the receivers. This includes effects such as path loss, shadowing, and fading. The latter is caused by multipath and dominates in indoor environments; see for example Fig.~\ref{fig:fading}.
        \end{bullets}%
        \blt[learn from data \ra DNN]The complexity of these phenomena calls for a PCD that learns to solve \eqref{eq:pcdprob} in a data-driven fashion. To this end, in this work, $\DecPCD$ is implemented using a DNN, as described next.

    \end{bullets}%
\end{bullets}%

\subsubsection{Architecture}
\label{sec:architecture_poschange}
\begin{bullets}%
    \blt[test statistic]Following standard practice, the detector is designed to  decide $\HoPCD$ when a
    detection statistic $\NetWei:\rfield^{\NumFea} \times
        \rfield^{\NumFea} \rightarrow \rfield$ exceeds a predefined threshold, and $\HzPCD$ otherwise. In this way, the problem of designing $\DecPCD$ becomes that of designing a function $\NetWei$.

    \blt[commutativity]
    \begin{bullets}%
        \blt[motivation]In principle, this function could be directly implemented
        as a DNN. However, such a simple approach would result in a non-commutative
        $\NetWei$, that is,  $\NetWei(\EstVecRss, \EstVecRss')$ will generally differ from
        $ \NetWei(\EstVecRss', \EstVecRss)$, which is clearly undesirable.
        To remedy this issue, a symmetrization technique will be adopted.
        \blt[technique]Specifically, $\NetWei$ will be implemented based on an auxiliary function $\AuxNetWei$ by setting $\NetWei(\EstVecRss, \EstVecRss') \define (\AuxNetWei(\EstVecRss, \EstVecRss') +
            \AuxNetWei(\EstVecRss', \EstVecRss)) / 2$. Observe that this implies that $\NetWei(\EstVecRss, \EstVecRss')=\NetWei(\EstVecRss', \EstVecRss)$ regardless of $\AuxNetWei$. Thereby, $\AuxNetWei$ can be safely implemented as a DNN.
    \end{bullets}%

    \blt[arch of subnet]The architecture of the subnetwork $\AuxNetWei$ is detailed next. Since $\AuxNetWei$ is concerned with dissimilarities, it is natural to include an
    initial non-trainable  layer that yields $10\log_{10}[\EstVecRss, \EstVecRss',
            \EstVecRss - \EstVecRss']$ when the input to the network is $[\EstVecRss, \EstVecRss']$. This facilitates learning from moderate-sized datasets. This layer is followed by  three hidden fully-connected layers with $512$ neurons and leaky
    ReLU activations~\cite{goodfellow2016}. The output
    layer contains a single neuron with a linear activation.

\end{bullets}%

\subsubsection{Data set}
\label{sec:dataset_poschange}
\begin{bullets}
    \blt[Overview]To train $\NetWei$,  a dataset comprising pairs of vectors from $\Dat$ is constructed. The feature vectors in $\NumPai$ of these pairs correspond to the same transmitter location. In the remaining $\NumPai$ pairs, they correspond to different transmitter locations.

    \blt[Data]
    \begin{bullets}
        \blt[Same location]Specifically, the pairs of the first kind are generated for
        $\IndPai=1,\ldots,\NumPai$ by first drawing $\IndDat\NotDat\IndPai$  uniformly at random
        from the set $\{1,\ldots,\NumDat\}$. Then, $\IndEst\NotDat\IndPai$ and
        $\IndEst\NotDat\IndPai'$ are drawn uniformly at random without replacement
        from $\{1,\ldots,\NumEst\}$. This  process results in the set
        $\DatSam\define\{(\EstVecRss\NotEst{\IndEst\NotDat\IndPai}(\Loc\NotDat{\IndDat\NotDat\IndPai}),
            \EstVecRss\NotEst{\IndEst'\NotDat\IndPai}(\Loc\NotDat{\IndDat\NotDat{\IndPai}})),~\IndPai=1,\ldots,\NumPai\}\subset\rfield^{\NumFea}\times\rfield^{\NumFea}$.
        \blt[Different location]To generate the  pairs
        of the second kind,
        draw $\IndDat\NotDat\IndPai$ and
        $\IndDat\NotDat\IndPai'$ uniformly at random  without replacement
        from the set $\{1,\ldots,\NumDat\}$ for
        $\IndPai=1,\ldots,\NumPai$. Drawing $\IndEst\NotDat\IndPai$ and
        $\IndEst\NotDat\IndPai'$ as before yields
        $\DatDif\define\{(\EstVecRss\NotEst{\IndEst\NotDat\IndPai}(\Loc\NotDat{\IndDat\NotDat\IndPai}),
            \EstVecRss\NotEst{\IndEst'\NotDat\IndPai}(\Loc\NotDat{\IndDat\NotDat{\IndPai'}})),~\IndPai=1,\ldots,\NumPai\}\subset\rfield^{\NumFea}\times\rfield^{\NumFea}$.
    \end{bullets}
    The DNN can then be trained on the dataset $\DatSam\cup\DatDif$.

    \blt[generation] Recall from Sec.~\ref{sec:problem} that obtaining $\Dat$ involves collecting the
    $\NumEst$ RSS estimates
    $\EstVecRss\NotEst{\IndEst}(\Loc\NotDat\IndDat)$, $
        \IndEst=1,\ldots,\NumEst$,
    for each of the $\NumDat$ locations $\Loc\NotDat\IndDat,~\IndDat=1,\ldots,\NumDat$. A simpler approach may be to collect a single estimate with a large $\NumSam$ so that it approximately equals
    $\VecRss(\Loc\NotDat\IndDat)$ and then generate the $\EstVecRss\NotEst{\IndEst}(\Loc\NotDat\IndDat)$ synthetically. The procedure is described next for the case where $\RxSig\NotFea{\IndFea}(\Loc,t)$ is approximately Gaussian distributed, which would be the case e.g. if $\TxSig(t)$ is an orthogonal frequency division multiplexing (OFDM) signal; see e.g. \cite{romero2013wideband}.

    To this end, express $\RxSig\NotFea{\IndFea}(\Loc,t)$ as
    $\RxSig\NotFea{\IndFea}(\Loc,t) =
        \Rss\NotFea{\IndFea}(\Loc)\samp\NotFea{\IndFea}(\Loc,t)$, where $\Rss\NotFea{\IndFea}(\Loc)$ is the true RSS and
    $\samp\NotFea{\IndFea}(\Loc,t)$ is a circularly symmetric zero-mean Gaussian random variable with
    unit variance uncorrelated over $t$. Then, \eqref{eq:rssestimator} becomes
    \begin{salign}
        \EstRss\NotFea{\IndFea}(\Loc,t) &
        = \frac{|\Rss\NotFea{\IndFea}(\Loc)|^2}{\NumSam} \sum_{\IndSam=1}^{\NumSam} |\samp\NotFea{\IndFea}(\Loc,t+\IndSam T)|^2                                                                                 \\\nonumber
        & = \frac{|\Rss\NotFea{\IndFea}(\Loc)|^2}{2\NumSam} \sum_{\IndSam=1}^{\NumSam} \left(\left[\sqrt{2}\text{Re}\{\samp\NotFea{\IndFea}(\Loc,t+\IndSam T)\}\right]\right.^2 \\
        &  + \left[\sqrt{2}\text{Im}\{\samp\NotFea{\IndFea}(\Loc,t+\IndSam T)\}\right]^2\Big).
    \end{salign}
    It follows that $2\NumSam\EstRss\NotFea{\IndFea}(\Loc,t)/|\Rss\NotFea{\IndFea}(\Loc)|^2$
    is a $\chi^2$ random variable with $2\NumSam$ degrees of freedom. Thus, $\NumEst$ estimates $\EstRss\NotEst\IndEst\NotFea{\IndFea}(\Loc,t)$, $\IndEst=1,\ldots,\NumEst$, can be obtained  by generating $\NumEst$ realizations of such a $\chi^2$ random variable.
\end{bullets}

\subsubsection{Training}
\label{sec:training_poschange}
The DNN is trained using a binary cross-entropy loss function. Within the
dataset, a subset comprising $\NumDat\Val$ validation points is
reserved for  validation, while the remainder $\NumDat\Tra = \NumDat
    - \NumDat\Val$ are used for training.
\subsection{Graph Neural Network based Spoofing Detection}
\label{sec:gnn_detection}

\begin{figure}[t]
    \centering
    \includegraphics[width=1.\columnwidth]{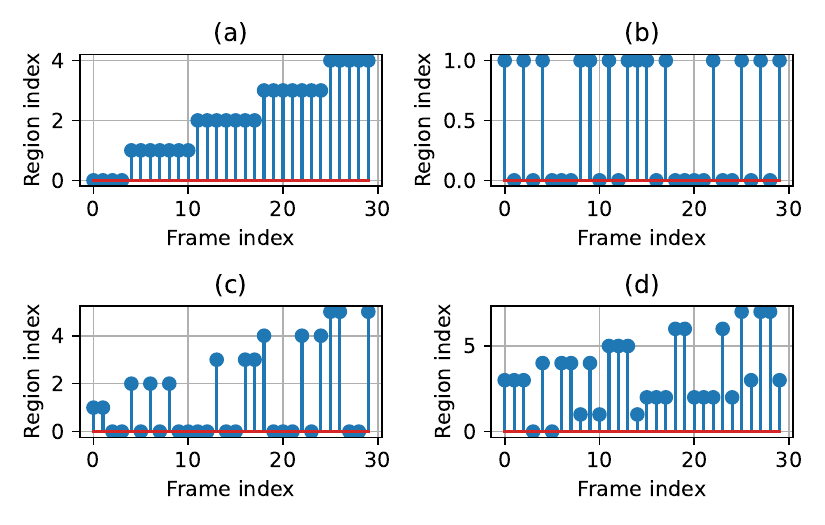}
    \caption{Examples of region sequences without an attack ((a)) and with an attack ((b)-(d)).}
    \label{fig:regions}
\end{figure}

\begin{bullets}%
    \blt[Overview]The decisions of the PCD for all pairs of frames will be used next to detect spoofing attacks.
    \blt[Intuition]
    \begin{bullets}%
        \blt[PCD \ra changes in propagation]To intuitively understand why this is possible, recall from Sec.~\ref{sec:poschange} that  the PCD decides $\HoPCD$ when the dissimilarity between the given feature vectors $\VecRss\NotFra{\IndFra}$ and $\VecRss\NotFra{\IndFra'}$ owes to the difference between the propagation phenomena experienced at $\Loc\NotFra{\IndFra}$ and $\Loc\NotFra{\IndFra'}$. In other words, if $\Loc\NotFra{\IndFra}$ and $\Loc\NotFra{\IndFra'}$ are so close that both points see similar propagation conditions to all receivers, the PCD decides $\HzPCD$.
        \blt[regions of constant propagation]For didactical purposes, it is useful to split the space into propagation \emph{regions} and assume that $\DecPCD(\VecRss\NotFra{\IndFra},\VecRss\NotFra{\IndFra'})=\HzPCD$ if $\Loc\NotFra{\IndFra}$ and $\Loc\NotFra{\IndFra'}$ belong to the same  region and  $\DecPCD(\VecRss\NotFra{\IndFra},\VecRss\NotFra{\IndFra'})=\HoPCD$ otherwise. Using the decisions of the PCD, one can therefore assign each frame in the given sequence to a region.

        \blt[figure]This assignment is illustrated in Fig.~\ref{fig:regions}, which sheds light into why it is possible to solve \eqref{eq:problem} using the decisions of the PCD.
        \begin{bullets}%
            \blt[no attack]Fig.~\ref{fig:regions}a shows an example where all frames are generated by a single moving user. Due to Assumption 1, groups of consecutive frames are declared by the PCD to belong to the same region.
            \blt[attack, no movement]In turn, Fig.~\ref{fig:regions}b depicts the case where two transmissions are concurrently taking place from different regions, which indicates the presence of an attack.
            \blt[attack+movement]Finally, Fig.~\ref{fig:regions}c and Fig.~\ref{fig:regions}d respectively correspond to the case where one or both of the concurrently transmitting users move.
        \end{bullets}%
    \end{bullets}%

    \blt[transitivity] It is important to note that the above considerations are provided to develop intuition, but in practice do not hold exactly. In particular, the decisions of the PCD will not generally be transitive, that is, it may hold that  $\DecPCD(\VecRss\NotFra{\IndFra},\VecRss\NotFra{\IndFra'})=\HzPCD$ and $\DecPCD(\VecRss\NotFra{\IndFra'},\VecRss\NotFra{\IndFra''})=\HzPCD$ but $\DecPCD(\VecRss\NotFra{\IndFra},\VecRss\NotFra{\IndFra''})=\HoPCD$. However, this may approximately hold.
    \blt[graph]For this reason, it is useful to construct a graph  $\Gra$ where the $\IndFra$-th node corresponds to  $\VecRss\NotFra{\IndFra}$ and there is an edge between nodes $\IndFra$ and $\IndFra'$ if $\DecPCD(\VecRss\NotFra{\IndFra},\VecRss\NotFra{\IndFra'})=\HzPCD$. Clearly, since $\DecPCD$ is commutative (cf. Sec.~\ref{sec:architecture_poschange}), this graph is undirected.

    \blt[motivation GNN]Clearly, if the space could be split into propagation regions, as discussed earlier, then this graph could be partitioned into one component per region and the presence of an attack would be characterized by an alternating pattern between components as in Fig.~\ref{fig:regions}b-\ref{fig:regions}d. However, since this is not exactly the case, it makes sense to train a GNN to detect attacks based on  $\Gra$ in a data-driven fashion.


\end{bullets}%



\subsubsection{Architecture}
\label{sec:architecture_gnn}
\cmt{General GNN}
\begin{bullets}
    \blt[overview]A GNN~\cite{scarselli2009graph} exploits the relation between node features and the graph topology by performing a sequence of message-passing steps or \emph{layers}, where the features associated with a node at layer $\LayerInd$ depend on the features of
    that node and
    the neighboring nodes at layer $\LayerInd-1$.
    \blt[Message Passing]Specifically, if $\bm \phi_{\NodeInd}^{(\LayerInd)} $ represents the features of node $\NodeInd$ at layer $\LayerInd$, then
    \begin{align}
        \bm \phi_{\NodeInd}^{(\LayerInd)} = G_1^{(\LayerInd)}\left(\bm \phi_{\NodeInd}^{(\LayerInd-1)}, \bigoplus_{\NodeInd'\in\Nei_{\NodeInd}} G_2^{(\LayerInd)}(\bm \phi_{\NodeInd}^{(\LayerInd-1)},\bm \phi_{\NodeInd'}^{(\LayerInd-1)})\right),
    \end{align}
    where $G_1^{(\LayerInd)}$ and $G_2^{(\LayerInd)}$ are conventional DNNs, $\Nei_{\NodeInd}$ contains the set of neighbors of node $\NodeInd$, and $\bigoplus$ is an aggregation operator such as a summation or maximum operator.
    \blt[output]The output of the GNN can be computed by another aggregation operator applied to the concatenation of the vectors $\bm \phi_{\NodeInd}^{(\NumLayer)}$ for all $\NodeInd$, where $\NumLayer$ is the number of layers.

\end{bullets}

\cmt{GNN spoofing detection}For the problem at hand, a test statistic is obtained with a GNN and then compared to a threshold to decide between $\hz$ and $\ho$. In the adopted architecture,
$\NumLayer=3$ layers and functions $G_1^{(\LayerInd)}$ and $G_2^{(\LayerInd)}$ are implemented as single-layer fully-connected DNNs with $64$ output neurons and ReLU activations. The operator $\bigoplus$ is a summation whereas the output of the GNN is obtained by averaging the features of all nodes at the last layer and applying a trainable affine transformation. Since the order of the nodes is relevant (cf. Fig.~\ref{fig:regions}), the features in the first layer are set so that $\bm \phi_{\NodeInd}^{(1)}$ equals the index of node~$\NodeInd$.


\subsubsection{Dataset}
\label{sec:dataset_gnn}

\begin{bullets}%
    \blt[frame sequence]
    To train the GNN, realizations of $\Gra$ must be generated under both $\hz$ and $\ho$. This involves generating frame sequences $\EstMatRss$ under both hypotheses.
    \begin{bullets}%
        \blt[H0]Under $\HzSD$, the trajectory $\Loc(t)$  of the (single) user is
        generated as follows. First, obtain the time duration of the frame sequence,
        given by $\NumFra\RatFra$, where $\RatFra$ is the number of frames per second. The length $\Delta\Loc$ of the
        trajectory is therefore $\Delta\Loc=\NumFra\RatFra\Spe$, where $\Spe$ is the speed of the user. Then, randomly draw a straight line segment of length $\Delta\Loc$ in $\Reg$. The trajectory is therefore
        $\Loc(t)=\LocLin + \VecDir \Spe t$, where $\LocLin$ is the starting point and $\VecDir$ is the unit direction vector of the line. For each $\IndFra$,
        $\EstVecRss[\IndFra]$ is  obtained by randomly selecting one of the vectors in $\Dat$ that correspond to  the
        location that lies closest to $\Loc(\IndFra/\RatFra)$.

        \blt[H1] Under $\HoSD$, the frame sequences of both users are generated following the above procedure. Then, the frame sequence
        $\EstVecRss_1[0],\ldots,\EstVecRss_1[\NumFra-1]$ of user-1 is merged with the
        frame sequence $\EstVecRss_2[0],\ldots,\EstVecRss_2[\NumFra-1]$ of user-2 into a
        sequence $\EstVecRss[0],\ldots,\EstVecRss[\NumFra-1]$ where either
        $\EstVecRss[\IndFra]=\EstVecRss_1[\IndFra]$  or
        $\EstVecRss[\IndFra]=\EstVecRss_2[\IndFra]$, both with probability 1/2 and independently
        along $\IndFra$.
    \end{bullets}%

\end{bullets}%




\subsubsection{Training}
\label{sec:training_gnn}
The GNN model is trained using a binary cross-entropy loss function.

\section{Performance Evaluation}
\label{sec:performance}
\cmt{Overview}

\cmt{Simulation setup}
\begin{bullets}
    \blt[Dataset]This section assesses the performance of the proposed scheme using the dataset from
    \cite{lohan2017crowdsourced}, which  contains  RSS measurements of $992$ WiFi access points at  $4846$ locations across 4 floors.
    \blt[Data preprocessing]To ensure a sufficient spatial density,  only the  measurements collected at $\NumDat=648$ locations on the first
    floor are used. $20\%$ of them are reserved for testing.
    %
    \blt[Samples]Since each measurement location lies out of the range of most of the access points, the $\NumFea=5$ access points that are measured at the greatest number of the selected locations are considered. The procedure described in Sec.~\ref{sec:dataset_poschange} is then used to generate $\NumEst=1000$ feature vectors for each location. A link to the code is provided on the first page.


    \blt[Benchmarks]
    The proposed algorithm, referred to as \emph{GNN-based Spoofing Detection} (GSD), is compared against four
    benchmarks which, along the lines of \cite{chen2007detecting,hoang2018soft,sobehy2020csi}, rely on clustering the feature vectors. This is intuitive as the frames transmitted from the same location will tend to be clustered together. The number of clusters is then used as a test statistic and the threshold is obtained to attain a target probability of false alarm ($\pfa$). The considered clustering algorithms include  \emph{density based spatial clustering
        of applications with noise} (DBSCAN) \cite{ester1996density}, \emph{hierarchical DBSCAN} (HDBSCAN) \cite{campello2013density},
    \emph{ordering
        points to identify the clustering structure} (OPTICS)
    \cite{ankerst1999optics}, and \emph{balanced iterative reducing and clustering
        using hierarchies} (BIRCH)~\cite{zhang1996birch}.
\end{bullets}

\cmt{Description of the experiments}
\begin{bullets}
    \blt[Experiment 1 - ROC] Fig.~\ref{fig:roc} depicts the receiver
    operating characteristic (ROC) curves \cite[Ch. 3]{kay2} of GSD and the benchmarks. It is seen that GSD results in a significantly higher probability of detection ($\pd$) for each $\pfa$.
    \begin{figure}[t]
        \centering
        \includegraphics[width=1.\columnwidth]{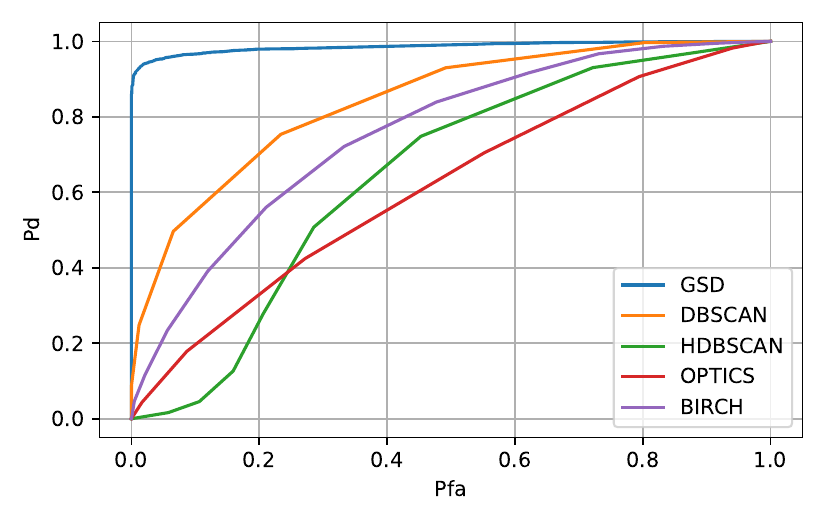}
        \caption{ROC curves of the proposed algorithm and the benchmarks (10
            frames/s, $\NumFra=30$, $\NumSam=150$, $\NumFea=5$).}
        \label{fig:roc}
    \end{figure}

    \blt[Experiment 2 - Pd vs speeds] Fig.
    \ref{fig:pd_speed} analyzes the influence of the speed of the users in the $\pd$ for a given $\pfa=0.1$ for the compared algorithms. Interestingly, speed seems to positively impact the $\pd$ of all algorithms, especially those based on clustering. This is because the
    the number of clusters per user increases with the speed and, therefore, the test statistic will tend to be more different between hypotheses.
    \begin{figure}[t]
        \centering
        \includegraphics[width=1.\columnwidth]{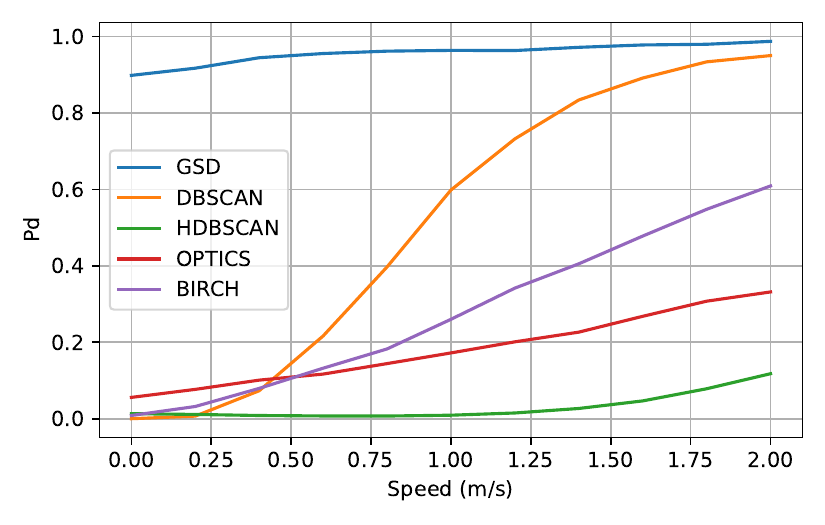}
        \caption{Pd vs speed of the proposed algorithm and the benchmarks for a
            fixed probability of false alarm $\pfa$ (10 frames/s, $\NumFra=30$,
            $\NumSam=150$, $\NumFea=5$, $\pfa=0.1$).
        }
        \label{fig:pd_speed}
    \end{figure}

    \blt[Experiment 3 - Pd vs number of frames] Fig. \ref{fig:pd_numframes}
    investigates the impact of $\NumFra$ on the detection performance. As expected,
    $\pd$ tends to increase with $\NumFra$. However, a wiggling effect is observed for the benchmarks. This does not vanish even if the number of Monte Carlo iterations is increased. The cause is the discrete nature of the test statistic of the benchmarks.
    \begin{figure}[t]
        \centering
        \includegraphics[width=1.\columnwidth]{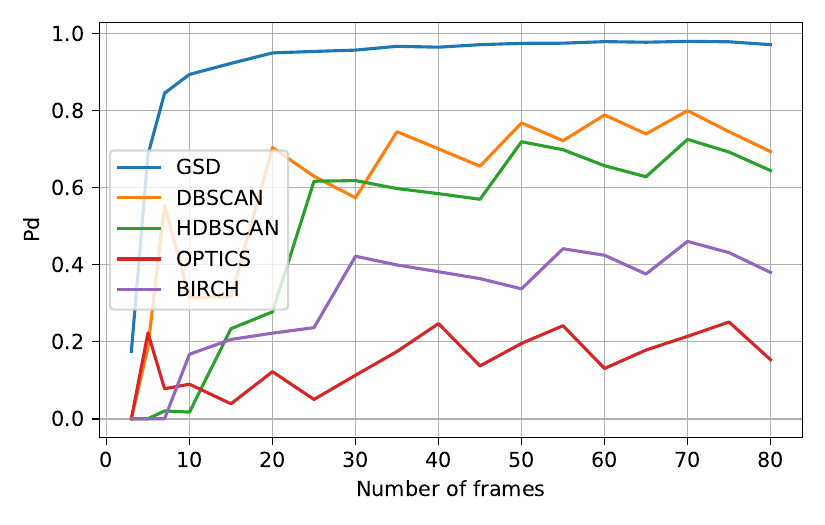}
        \caption{Pd vs number of frames of the proposed algorithm and the
            benchmarks for a fixed probability of false alarm $\pfa$ (10 frames/s,
            $\NumSam=150$, $\NumFea=5$, $\pfa=0.1$).}
        \label{fig:pd_numframes}
    \end{figure}

    \blt[Experiment 4 - PD vs num samples] Finally,
    Fig. \ref{fig:pd_num_samples} analyzes how $\pd$ evolves as a function of $\NumSam$. It is remarkable that GSD attains a very large $\pd$ even for a small $\NumSam$, which suggests that the PCD successfully learned to distinguish the two sources of variability in the feature vectors described in Sec.~\ref{sec:poschange}.
    \begin{figure}[t]
        \centering
        \includegraphics[width=1.\columnwidth]{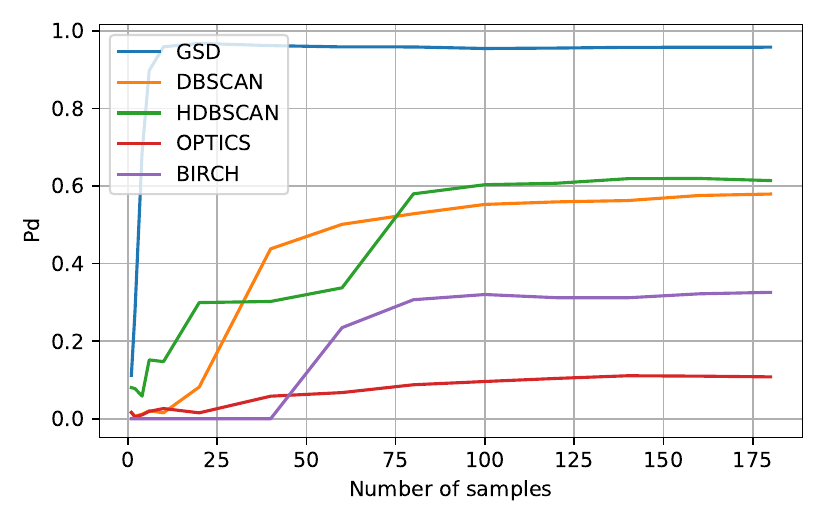}
        \caption{$\pd$ vs number of frames of the proposed algorithm and the
            benchmarks for a fixed probability of false alarm $\pfa$ (10 frames/s,
            $\NumFra=30$, $\NumFea=5$, $\pfa=0.1$).}
        \label{fig:pd_num_samples}
    \end{figure}

\end{bullets}

\section{Conclusion}
\label{sec:conclusion}
\cmt{Summary of the paper} This work considered the problem of detecting spoofing attacks in the physical layer, which is motivated by the vulnerability of cryptographic techniques when the credentials of the legitimate user are compromised. Unfortunately, prior schemes based on RSS measurements raise false alarms in the presence of channel changes or user movement. To remedy this limitation, this work
introduced a deep learning detector robust to these effects. Since localization-based approaches would suffer from a low spatial resolution due to multipath effects, a position change detector based on a deep neural network is used to build a graph embedding of the RSS features. The temporal pattern of changes in the propagation conditions captured by this embedding is then learned by a GNN, which then decides on the presence of a spoofing attack.
Empirical evaluation  with real-world data
showcases the effectiveness of this scheme as well as its robustness to the mobility of the user and attacker.


\printmybibliography

\end{document}